\documentstyle[preprint, aps]{revtex}
\draft
\tighten

\begin{document}
\title{Charge and stability of strange quark matter at finite temperature}
\author{Yun Zhang$^{1,3}$ and Ru-Keng Su$^{2,3}$}
\address{$^1$Surface Physics Laboratory (National Key Laboratory), Fudan University,\\
Shanghai 200433 , P. R. China \\
$^2$China Center of advanced Science and Technology (World Laboratory)\\
P. O. Box 8730, Beijing 100080, P. R. China \\
$^3$Department of physics, Fudan University, Shanghai 200433 , P. R. China}
\maketitle

\begin{abstract}
By means of the quark mass density- and temperature- dependent model, it is
found that the negative charge and the higher strangeness fraction are in
favor of the stability of strange quark matter at finite temperature. A
critical baryon number density $n_{Bc}$ has been found. The strange quark
matter is unstable when $n_B<$ $n_{Bc}$. The charge and strangeness fraction
dependence of stability for strange quark matter is addressed.
\end{abstract}


\section{Introduction}

It has been found from the theoretical study that the introduction of
strange quarks into a plasma with two flavors could lower the Fermi energy
of the system and thus the mass of quark matter \cite{1}. If its mass is
lower than the mass of hyperonic matter with the same strangeness fraction,
or even lower than the non-strange nucleonic matter, the strange quark
matter (SQM)\ would be the true ground state of QCD. Therefore, to search
the SQM\ in the laboratory is an important task for recent heavy ion
collision experiments at ultrarelativistic energy \cite{su2}.

The essential problem for detectability of SQM in heavy ion experiments is
to study its stability during the formation of quark gluon plasma (QGP).
Many properties of stability of SQM, for example, the ''stability window'' 
\cite{su3,su4,su5}, the charge-dependence \cite{su6,su7,su8,su9} and the
strangeness fraction-dependence \cite{su10,su11,su12,su13,su14} of stability
have been widely discussed by many authors. They employed different models
to investigate this problem. But unfortunately, many conclusions are
different, in particular, part of them, say, the charge dependence is
contrary.

The electric charge is of vital importance to the experimental searches of
SQM. By means of MIT bag model, Jaffe and his co-workers \cite{su3,su6,su7}
suggested that the SQM\ is slightly positively electric charged. They
considered strong and weak decay by nucleon and hyperon emission together
and concluded that the stable SQM will have a low but positive charge to
mass ratio. Using the same model as that of Jaffe et al. but considering the
initial condition of possible strangelet production in relativistic heavy
ion collision carefully, Greiner and his co-workers \cite{su7,su12} argued
that strangelets are most likely highly negatively charged. They also
emphasized that the stability of infinite SQM depends on the bag constant $B$%
. An infinite SQM, treated as a gas of noninteracting quarks, is absolutely
stable for $B^{1/4}\simeq 145\mbox{MeV}$, and metastable for $B^{1/4}\simeq
150-200\mbox{MeV}$, but unstable for larger bag constants. Of course, this
conflict of conclusions is limited in the framework of MIT bag model and at
zero temperature.

Employing another model, namely, the quark mass density-dependent (QMDD)
model, Peng and his co-workers reconsidered this problem \cite{su9}. They
introduced a critical density and proved that SQM cannot maintain its flavor
equilibrium, and thus no longer be stable under this critical density. They
found that proper negative charges can lower the critical density and thus
make the SQM become more stable. Their conclusion is not surprised if one
notice the confinement mechanism of QMDD\ model. In QMDD model, the mass of $%
u,d$ and $s$ quarks are given by \cite{su16,su17,su18,su4,su9,su7,su3} 
\begin{eqnarray}
m_q &=&{\frac B{3n_B}},\hspace{0.8cm}(q=u,d,\bar{u},\bar{d}),  \label{su1} \\
m_{s,\bar{s}} &=&m_{s0}+{\frac B{3n_B}},  \label{su2}
\end{eqnarray}
where $n_B$ is the baryon number density, $m_{s0}$ is the current mass of
the strange quark matter and $B$ is the vacuum energy density inside bag
(bag constant). The mechanism of confinement can be mimicked through the
hypothesis Eqs.(\ref{su1}) and (\ref{su2}). The masses of quarks become
infinitely large as the volume increases to infinity or the density
decreases to zero. So that the vacuum is unable to support it \cite{su4,su18}%
. Noting that the boundary condition of confinement of MIT\ bag model
corresponds to the zero quark mass inside the bag but infinity at the
boundary or outside the bag, we see that the confinement mechanisms are
similar to these two models. Due to this correspondence, as pointed out by
Benrenuto and Lugones \cite{su4}, in almost all cases, the properties of
SQM\ given by QMDD\ model are nearly the same as those obtained in the MIT
bag model. The conclusion of Peng et al. supports the result given by
Greiner et al.

This paper evolves from an attempt to extend these discussions to finite
temperature. The reasons of this extension are: at first, the SQM might be
possible and formed during the phase transition from hadronic matter to a
deconfined QGP. From lattice QCD calculations, the phase transition of
deconfinement will happen at critical temperature. Secondly, as was pointed
out by our previous papers \cite{su5,su19}, QMDD model cannot be used to
explain the process of the quark deconfinement phase transition because the
quark confinement is permanent in this model. Instead of this model, a quark
mass density- and temperature-dependent (QMDTD)\ model in which the quark
confinement is not permanent has been suggested by us \cite{su5,su19}. As
its application, we hope to employ QMDTD\ model to discuss the charge and
strangeness fraction dependences of SQM at finite temperature.

The organization of this paper is as follows: In next section, we review the
QMDTD\ model and give the main formulae which are necessary for studying the
thermodynamical behaviors of SQM. In section 3, we give the result of
numerical calculations for QMDTD\ model and investigate the stability of hot
SQM in detail. Our conclusions are summarized in the last section.

\section{Quark mass density- and temperature-dependent models}

To overcome the difficulties of QMDD\ model which appear at finite
temperature, in refs.\cite{su5,su19}, based on the Friedberg-Lee model\cite
{su20}, we suggested a QMDTD model. The basic difference between our model
and QMDD\ model is that, instead of a constant $B$ in Eqs.(\ref{su1}) and (%
\ref{su2}), we argued that $B$ must be a function of temperature and choose $%
B(T)$ as

\begin{eqnarray}
\ B &=&B_0\left[ 1-\left( \frac T{T_c}\right) ^2\right] ,0\leq T\leq T_c
\label{su3} \\
B &=&0,\text{ }T>T_c,  \label{su4}
\end{eqnarray}
where $B_0$ is bag constant at zero temperature and $T_c$ is the critical
temperature of quarks deconfinement. Substituting Eq.(\ref{su3}) into Eqs.(%
\ref{su1}) and (\ref{su2}), we find the masses of $u,d,s$ quarks and their
corresponding anti-quarks as 
\begin{eqnarray}
m_q &=&{\frac{B_0}{3n_B}}\left[ 1-\left( \frac T{T_c}\right) ^2\right] ,%
\hspace{0.8cm}(q=u,d,\bar{u},\bar{d}),  \label{su5} \\
m_{s,\bar{s}} &=&m_{s0}+{\frac{B_0}{3n_B}}\left[ 1-\left( \frac T{T_c}%
\right) ^2\right] ,  \label{su6}
\end{eqnarray}
when $0\leq T\leq T_c$. When $T\geq T_c$, the masses of quarks are
independent of temperature: $m_{u,d,\bar{u},\bar{d}}=0$ and $m_{s,\bar{s}%
}=m_{s0}$. Due to the difference between Eqs.(\ref{su1}),(\ref{su2}) and
Eqs.(\ref{su5}),(\ref{su6}), as was proved in ref.\cite{su5}, we found that
when $n_B$ approaches to zero, the temperature $T$ $\rightarrow T_c$ in our
model, but diverges in QMDD model. In QMDTD\ model, the masses of $u,d$
quarks become zero (or constant for strange quarks) when the temperature
approaches $T_c$. This means that the quarks can be deconfined. Here, the
vacuum energy density $B$ is a decreasing function as temperature increases.
As was pointed out by Greiner et al., the chosen values of $B$ affect the
stability of SQM remarkably \cite{su8}. Therefore, it is of interest to
consider the effect of $B(T)$.

Treating the infinite SQM as a gas of noninteracting quarks and electrons 
\cite{su6,su9}, we find that the thermodynamical potential density of SQM is 
\begin{equation}
\Omega =\sum_i\Omega _i=-\sum_i\frac{g_iT}{(2\pi )^3}%
%TCIMACRO{\dint }
%BeginExpansion
\displaystyle \int %
%EndExpansion
_0^\infty dk{\frac{dN_i}{dk}}\ln \left( 1+e^{-\beta (\varepsilon _i(k)-\mu
_i)}\right) ,  \label{su7}
\end{equation}
where $i$ stands for $u,d,s$ (or $\bar{u},\bar{d},\bar{s}$ ) and the
electron $e$($e^{+}$), $g_i=6$ for quarks and antiquarks, $g_i=2$ for $e$
and $e^{+}$. $\varepsilon _i(k)=\sqrt{m_i^2+k^2}$ is the single particle
energy and $m_{i\text{ }}$is given by Eqs.(\ref{su5}),(\ref{su6}) for quarks
and antiquarks. $\mu _i$ is the chemical potential (for antiparticle $\mu _{%
\bar{i}}=-\mu _i$). The number density of each particle can be obtained from
eq.(\ref{su7}) by means of 
\begin{equation}
n_i=-{\frac 1V}\left. {\frac{\partial \Omega }{\partial \mu _i}}\right|
_{T,n_B},  \label{su8}
\end{equation}
and we find 
\begin{equation}
\Delta n_i=n_i-n_{\bar{i}}=\frac{g_i}{(2\pi )^3}%
%TCIMACRO{\dint }
%BeginExpansion
\displaystyle \int %
%EndExpansion
_0^\infty d^3k\left( \frac 1{\exp [\beta (\varepsilon _i-\mu _i)]+1}-\frac 1{%
\exp [\beta (\varepsilon _i+\mu _i)]+1}\right) .  \label{su9}
\end{equation}
For SQM, the baryon number density satisfies 
\begin{equation}
n_B={\frac 13}(\Delta n_u+\Delta n_d+\Delta n_s)\hspace{0in}.  \label{su10}
\end{equation}
The electric charge density $Q$ of SQM reads 
\begin{equation}
Q=\frac 23\Delta n_u-%
%TCIMACRO{\dfrac 13 }
%BeginExpansion
{\displaystyle {1 \over 3}}%
%EndExpansion
\Delta n_d\hspace{0in}-%
%TCIMACRO{\dfrac 13 }
%BeginExpansion
{\displaystyle {1 \over 3}}%
%EndExpansion
\Delta n_s-\Delta n_e,  \label{su11}
\end{equation}
and the light quarks are converted to strange quarks through the weak
processes \cite{su9,su17} 
\begin{equation}
u+d\longleftrightarrow u+s,s\rightarrow u+e^{-}+\bar{\nu}_e,d\rightarrow
u+e^{-}+\bar{\nu}_e,u+e^{-}\rightarrow d+\nu _e.  \label{su12}
\end{equation}
Neglecting the contribution of neutrinos \cite{su3,su4,su17,su18}, we obtain
the conditions of chemical equilibrium from the reactions (\ref{su12}) as 
\begin{eqnarray}
\mu _s &=&\mu _d,\hspace{0.5cm}  \label{su13} \\
\mu _s &=&\mu _u+\mu _e.  \label{su14}
\end{eqnarray}

Noting that the chemical potentials of particles and anti-particles are
determined by Fermi distribution Eq.(\ref{su9}), we see that Eqs.(\ref{su5}%
), (\ref{su6}), (\ref{su9}), (\ref{su10}), (\ref{su11}), (\ref{su13}) and (%
\ref{su14}) form an equation group and can be solved self-consistently. The
solution of this equation group determines the chemical stability of SQM.
The results of numerical calculation will show in the next section. Here we
want to point out that at a fixed temperature $T$ and for a fixed value of
charge density $Q$ this equation group which determines the configuration of
SQM has no solution under a critical baryon number density $n_{Bc}$. This
means that the SQM is unstable when $n_B<n_{Bc}$. With the values of $Q$
changed and the temperatures limited between $0\leq T\leq T_c$, the
condition $n_B\geq n_{Bc}$ determines a stable area which might be useful
for the detectability of SQM in experiments of high energy heavy ion
collisions.

\section{Conclusion and discussion}

The numerical calculations of the equation group mentioned above have been
done with the parameters set:

\begin{equation}
B_0=170\mbox{MeV}\mbox{fm}^{-3},m_{s0}=150\mbox{MeV},T_c=170\mbox{MeV}.
\label{o10}
\end{equation}

Firstly, we study a SQM system with neutrality charge $Q=0$, for a given $n_B
$, the fractions of different quarks are given by 
\begin{equation}
F_{q(q=u,d,s)}=%
%TCIMACRO{\dfrac{\Delta n_q}{3n_B}}
%BeginExpansion
{\displaystyle {\Delta n_q \over 3n_B}}%
%EndExpansion
.  \label{su15}
\end{equation}
At a fixed temperature $T=100\mbox{MeV}$, the $u,d,s$ quarks fraction are
drawn in Fig.1 by solid line, dashed line and dotted line, respectively. At
high baryon number density, all of the $u,d,s$ quarks tend to become a
triplicate. When the density becomes lower, $u$ fraction increases and $s$
fraction decreases monotonously. We see from this figure that the fractions
of $u,d,s$ quarks satisfy an inequality: 
\begin{equation}
F_d>F_u>F_s.  \label{su16}
\end{equation}
In particular, we would like to emphasize that, as shown in Fig.1, three
different quark fraction curves stop at a critical baryon number density ($%
n_{Bc}=8.28\times 10^{-3}\mbox{fm}^{-3}$). One can not find real roots for $%
F_d,F_u$ and $F_s$ provided $n_B<n_{Bc}$. At $n_{Bc}$, $F_s=0.107\neq 0$. In
ref.\cite{su9}, the authors defined the critical density by the condition $%
F_s=0$. Of course, the SQM\ can not maintain the chemical equilibrium if no
strange quarks exist. But as shown in Fig.1, this is only a sufficient but
is not the necessary condition. Because, one cannot find solution in the
ranges $0<F_s<0.107$.

To compare the QMDD model in which $B$ is a constant and our QMDTD model in
which $B$ satisfies Eqs.(\ref{su3}) and (\ref{su4}), we plot the critical
density $n_{Bc}$ vs temperature $T$ with neutral charge density $Q=0$ in
figure 2, where the dashed line and the solid line refer to QMDD\ model and
QMDTD\ model, respectively. We see that the curve given by QMDD\ model can
never be zero even the temperature increases to $T>T_c$. It can easily be
understood if we notice that $m_q$ approaches to infinity when $%
n_B\rightarrow 0$ from Eqs.(\ref{su1}) and (\ref{su2}). To excite a particle
with infinite mass must spend infinite energy and then infinite temperature.
Because of $n_{Bc}\neq 0$, when $n_B<n_{Bc}$, the system would still be
unstable even at very high temperature for QMDD\ model. But for QMDTD\
model, we see from the solid line that $n_{Bc}=0$ at $T=T_c=170\mbox{MeV}$.
It means that the SQM is completely stable and may be detected from high
energy heavy ion collision.

Now we are in a position to address the charge dependence of stability for
SQM. The critical density vs temperature curves with various charges are
shown in figure 3, where the solid line, dotted line and dashed line refer
to $Q=0,-0.1e\mbox{fm}^{-3},-0.2e\mbox{fm}^{-3}$, respectively. We see that
the critical density $n_{Bc}$ decreases rapidly and the stable area expands
widely when $Q$ decreases to more negative. So the negative charge is in
favor of the stability of SQM. Our result supports the conclusion given by
Greiner et al. \cite{su8} and extends their result to finite temperature.

To study the effect of the strangeness fraction, we show the same curves as
that of Fig.1 but for $Q=-0.2e\mbox{fm}^{-3}$ in figure 4. Comparing Fig.1
with Fig.4, we find that all of three curves tend to become a triplicate for 
$Q=-0.2e\mbox{fm}^{-3}$ at high density, too. And there are two major
differences between these two figures, firstly, the critical density
decreases from $8.28\times 10^{-3}\mbox{fm}^{-3}$ to $5.02\times 10^{-4}%
\mbox{fm}^{-3}$ when charge density $Q$ changes from neutrality to
negativity $-0.2e\mbox{fm}^{-3}$; secondly, Eq.(\ref{su16}) becomes

\begin{equation}
F_d>F_s>F_u.  \label{su17}
\end{equation}
The fraction of $s$ quarks is higher than the fraction of $u$ quarks. This
is of course reasonable because $s$ quark has negative charge and $u$ quark
positive. The critical density $n_{Bc}$ decreases as the $F_s$ increases.
The higher strangeness fraction is in favor of the stability of SQM. To
illustrate this result more transparently, using the expression of
strangeness fraction 
\begin{equation}
f_s=%
%TCIMACRO{\dfrac{\Delta n_s}{n_B}}
%BeginExpansion
{\displaystyle {\Delta n_s \over n_B}}%
%EndExpansion
,  \label{ofs}
\end{equation}
which is usually employed by many references, we draw the critical
strangeness fraction $f_{sc}$ vs temperature $T$ curves in figure 5, where
the solid line and dashed line refer to $Q=0$ and $-0.2e\mbox{fm}^{-3}$,
respectively. The $f_{sc}$ is strangeness fraction corresponding to the
critical density $n_{Bc}$ at the same conditions of charge and temperature.
We see from Fig. 5 the critical strangeness fraction increases with
increasing temperature. The stable area corresponds to the larger negative
charge and higher strangeness fraction.

Finally, we hope to investigate the stability of SQM from another view
point. Instead of considering the chemical equilibrium, we study the
behavior of chemical potentials for electron and $u$ quark, respectively.
The reason for this choice is that the electron ($u$ quark) is the particle
with minimum mass but negative (positive) charge in our model. The smaller
the particle mass is, the lower the chemical potential will be. When
chemical potential tends to be negative, the system becomes unstable because
the Fermi surface is negative in this case. Therefore, we can use the
conditions $\mu _e\geq 0$, $\mu _u\geq 0$ to determine the stability of SQM.

The chemical potential of electron $\mu _e$ vs baryon number density $n_B$
at a fixed temperature $T=100\mbox{MeV}$ with various positive charge
density $Q=1\times 10^{-4},1\times 10^{-3},3\times 10^{-3},5\times 10^{-3}$
and $7\times 10^{-3}e\mbox{fm}^{-3}$ are drawn in figure 6. We see that $\mu
_e$ becomes negative and the system becomes unstable when $Q<0.001e\mbox{fm}%
^{-3}$. Similarly, the chemical potential of $u$ quark $\mu _u$ vs baryon
number density $n_B$ at a fixed temperature $T=100\mbox{MeV}$ with various
negative charge density $Q=-0.2,-0.322,-0.4,$ $-0.6$ and $-0.8$ $e\mbox{fm}%
^{-3}$ are drawn in figure 7. We see from this figure that when $Q<$ $-0.322$%
, the curve will pass through a negative area and thus SQM\ is unstable.
Therefore, at temperature $T=100\mbox{MeV}$, the charge density must be
limited to $0.001e\mbox{fm}^{-3}>Q>-0.322e\mbox{fm}^{-3}$ for stable SQM.

\section{\protect\smallskip Summary}

In summary, by means of the QMDTD\ model, we find that the negative charge
and the higher strangeness fraction are in favor of the stability of SQM at
finite temperature. This result is in agreement with that given by Greiner
et al.\cite{su8} at zero temperature, but we extend their conclusion to
finite temperature. A critical baryon number density $n_{Bc}$ under which
the SQM is unstable has been obtained. We find that $n_{Bc}$ decreases
rapidly when charge density $Q$ becomes more negative and approaches to zero
at the deconfinement temperature $T_c$. By using the chemical potentials of
electron and $u$ quark, we suggest a method to determine the limit of charge
density for the stability of SQM. We hope the stable areas given by our
model can have impact on the detectability of SQM\ in future heavy ion
experiments.

\section{Acknowledgment}

This work was supported in part by the NNSF of China under contract Nos.
19975010, 10047005.

\section{Figure Captions}

Fig.1 The quark fraction $F_q$ as a function of baryon number density $n_B$
with neutral charge, $F_u,F_d$ and $F_s$ are represented by solid, dashed
and dotted lines, respectively. Here $F_d>F_u>F_s.$

Fig.2 The critical density $n_B$ as a function of temperature $T$ with
neutral charge, the dashed line is for QMDD\ model and the solid line is for
QMDTD\ model.

Fig.3 The critical density $n_B$ as a function of temperature $T$ with
various charge density $Q=0,-0.1$ and $-0.2e\mbox{fm}^{-3}$, they are
represented by solid, dotted and dashed lines, respectively.

Fig.4 The quark fraction $F_q$ as a function of baryon number density $n_B$
with a negative charge density $Q=-0.2e\mbox{fm}^{-3}$, $F_u,F_d$ and $F_s$
are represented by solid, dashed and dotted lines, respectively. Now $%
F_d>F_s>F_u.$

Fig.5 The critical strangeness fraction $f_{s\text{ }}$as a function of
temperature $T$. The solid line is for $Q=0$ and the dashed line is for $%
Q=-0.2e\mbox{fm}^{-3}$.

Fig.6 The chemical potential of electron $\mu _e$ as a function of baryon
number density $n_B$ with the positive charge density $Q=1\times
10^{-4},1\times 10^{-3},3\times 10^{-3},5\times 10^{-3}$ and $7\times
10^{-3}e\mbox{fm}^{-3}$. At a fixed temperature $T=100\mbox{MeV}$, they are
represented by solid lines respectively.

Fig.7 The chemical potential of $u$ quark $\mu _u$ as a function of baryon
number density $n_B$ with the negative charge density $%
Q=-0.2,-0.322,-0.4,-0.6,$and $-0.8e\mbox{fm}^{-3}$. At a fixed temperature $%
T=100\mbox{MeV}$, they are represented by solid lines respectively.


\begin{references}
\bibitem{1}  E. Witten, Phys. Rev. {\bf D30,} 272 (1984).

\bibitem{su2}  J. Barrette, et al., Phys. Lett. {\bf B252}, 550 (1990), M.
Aoki, et al., Phys. Rev. Lett. {\bf 69}, 2345 (1992), K. Borer, et al.,
Phys. Rev. Lett. {\bf 72}, 1415 (1994), D. Beavis, et al., Phys. Rev. Lett. 
{\bf 75}, 3078 (1995), T. Armstong, et al., Phys. Rev. Lett. {\bf 79}, 3612
(1997), G.Appleqvist, et al., Phys. Rev. Lett. {\bf 76}, 3907 (1996).

\bibitem{su3}  E. Farhi and R. L. Jaffe, Phys. Rev. {\bf D30,} 2379 (1984).

\bibitem{su4}  O. G. Benrenuto and G. Lugones, Phy. Rev. {\bf D51}, 1989
(1995), G. Lugones and O. G. Benrenuto, Phy. Rev. {\bf D52}, 1276 (1995).

\bibitem{su5}  Y. Zhang and R. K. Su, Phys. Rev. {\bf C65}, 035202 (2001).

\bibitem{su6}  M. S. Berger and R. L. Jaffe, Phys. Rev. {\bf C35}, 213
(1987).

\bibitem{su7}  E. P. Gilson and R. L. Jaffe, Phys. Rev. Lett. {\bf 71}, 332
(1993).

\bibitem{su8}  J. Schaffner-Bielich, C. Greiner, A. Diener and H.
St\"{o}cker, Phys. Rev. {\bf C55,} 3038{\bf \ }(1997).

\bibitem{su9}  G. X. Peng, H. C. Chiang, P.Z. Ning and B. S. Zou, Phys. Rev. 
{\bf C59}, 3452 (1999).

\bibitem{su10}  C. Greiner, P. Koch and H. St\"{o}cker, Phys. Rev. Lett {\bf %
58, }1825{\bf \ (}1987), C. Greiner and H. St\"{o}cker, Phys. Rev. {\bf D44,}
3517 (1991).

\bibitem{su11}  J. Scaffner, C. B. Dover, A. Gal, D. J. Millener, C. Greiner
and H. St\"{o}cker, Ann. Phys. (N. Y.) {\bf 235}, 35 (1994).

\bibitem{su12}  C. Greiner, J. Phys. G:\ Nucl. Part. Phys. {\bf 25}, 389
(1999).

\bibitem{su13}  P. Wang, R. K. Su, H. Q. Song and L. L. Zhang, Nucl. Phys. 
{\bf A653}, 166 (1999).

\bibitem{su14}  L. L. Zhang, H. Q. Song, P. Wang and R. K. Su, J. Phys. G:\
Nucl. Part. Phys. {\bf 26}, 1301 (2000).

\bibitem{su16}  G. N. Fowler, S. Raha and R. M. Weiner, Z. Phys. {\bf C9},
271 (1981).

\bibitem{su17}  S. Chakrabarty, Phys. Rev. {\bf D43}, 627 (1991), ibid {\bf %
48}, 1409 (1993).

\bibitem{su18}  S. Chakrabarty, S. Raha and B. Sinha, Phys. Lett. {\bf B229}%
, 112 (1989).

\bibitem{su19}  Y. Zhang, R. K. Su, S. Q. Ying and P. Wang, Europhys. Lett 
{\bf 56}, 361 (2001).

\bibitem{su20}  T. D. Lee, Particle Physics and Introduction to Field Theory
(Harwood Academic, Chur, 1981).
\end{references}
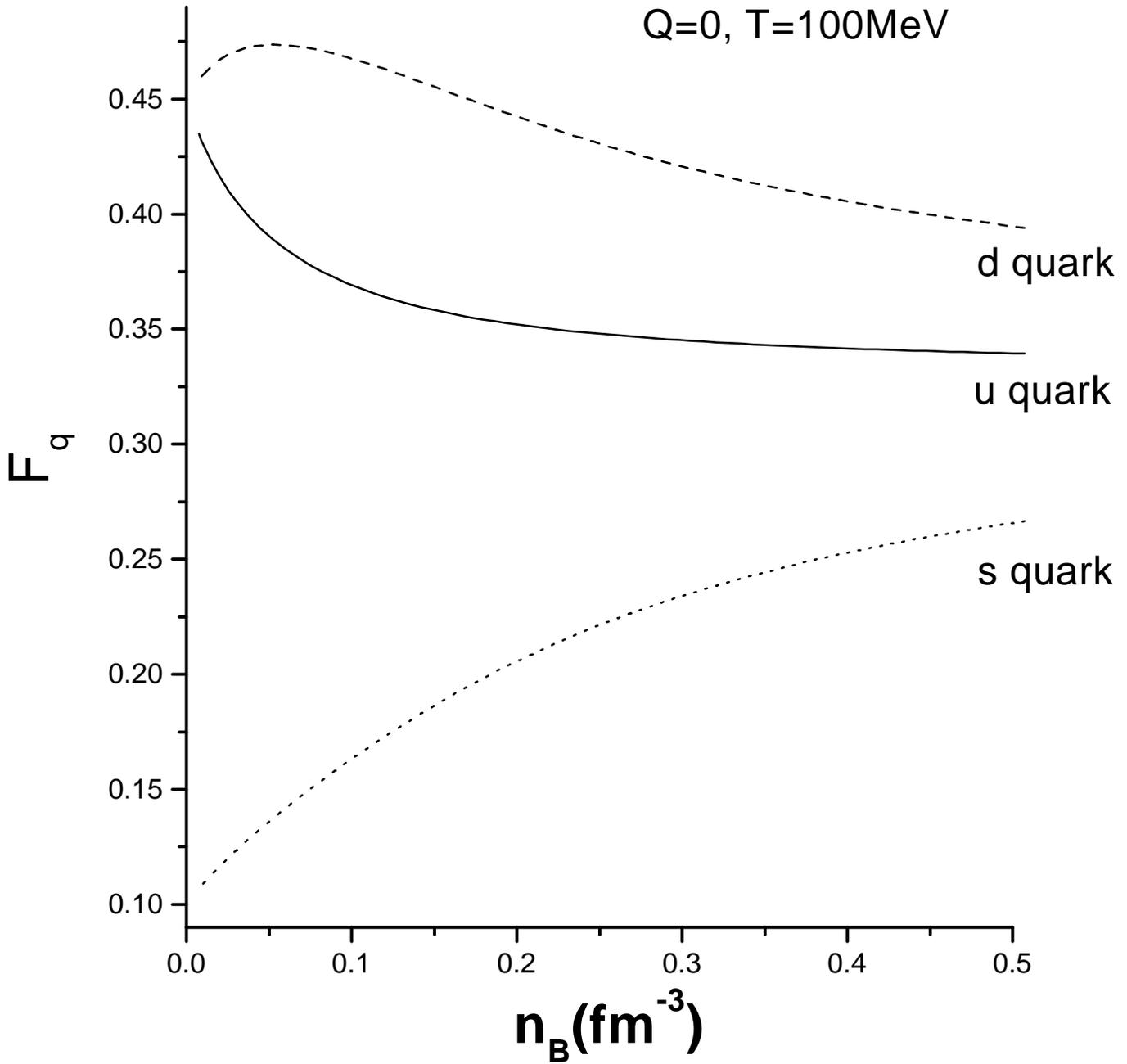
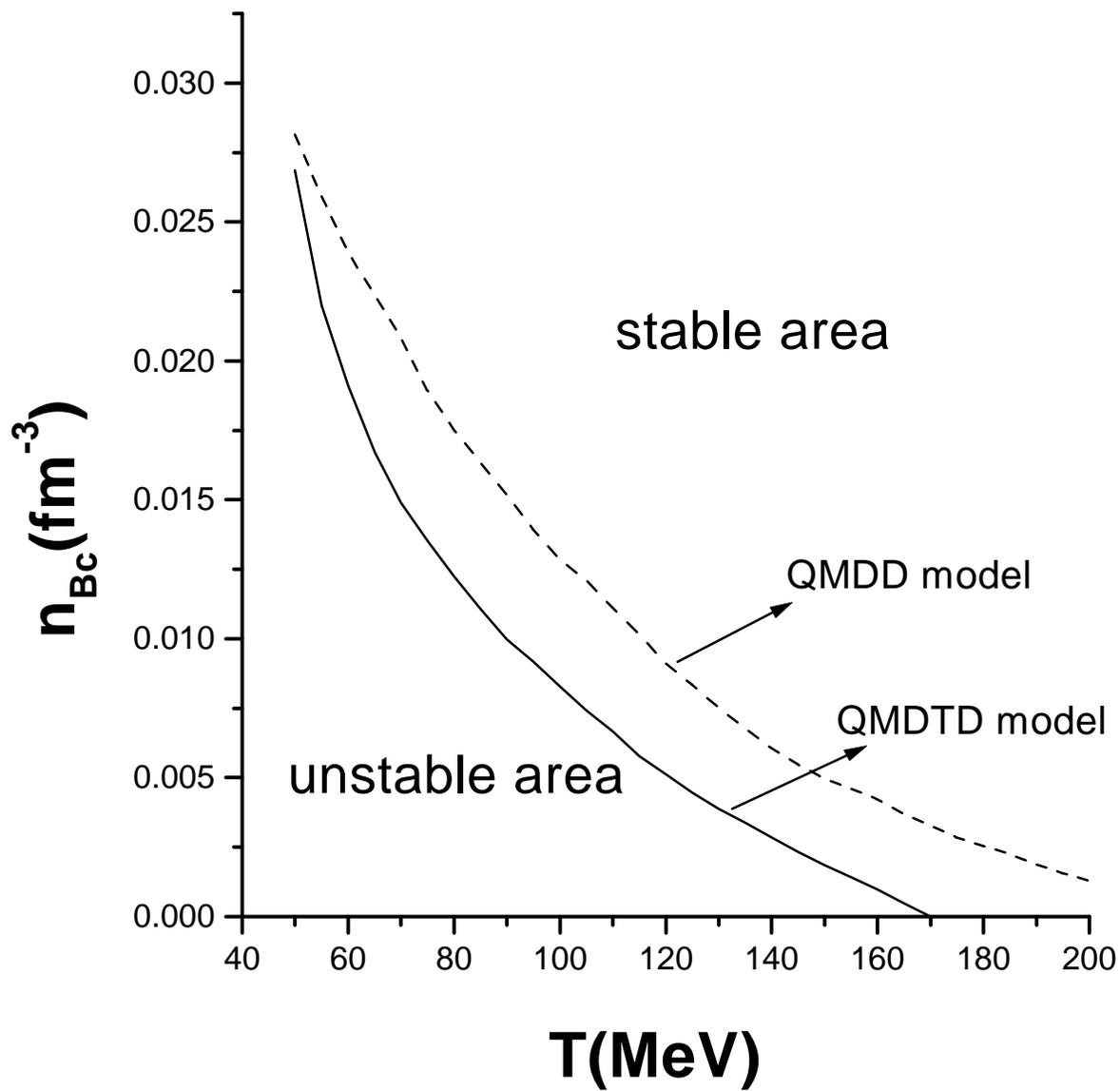
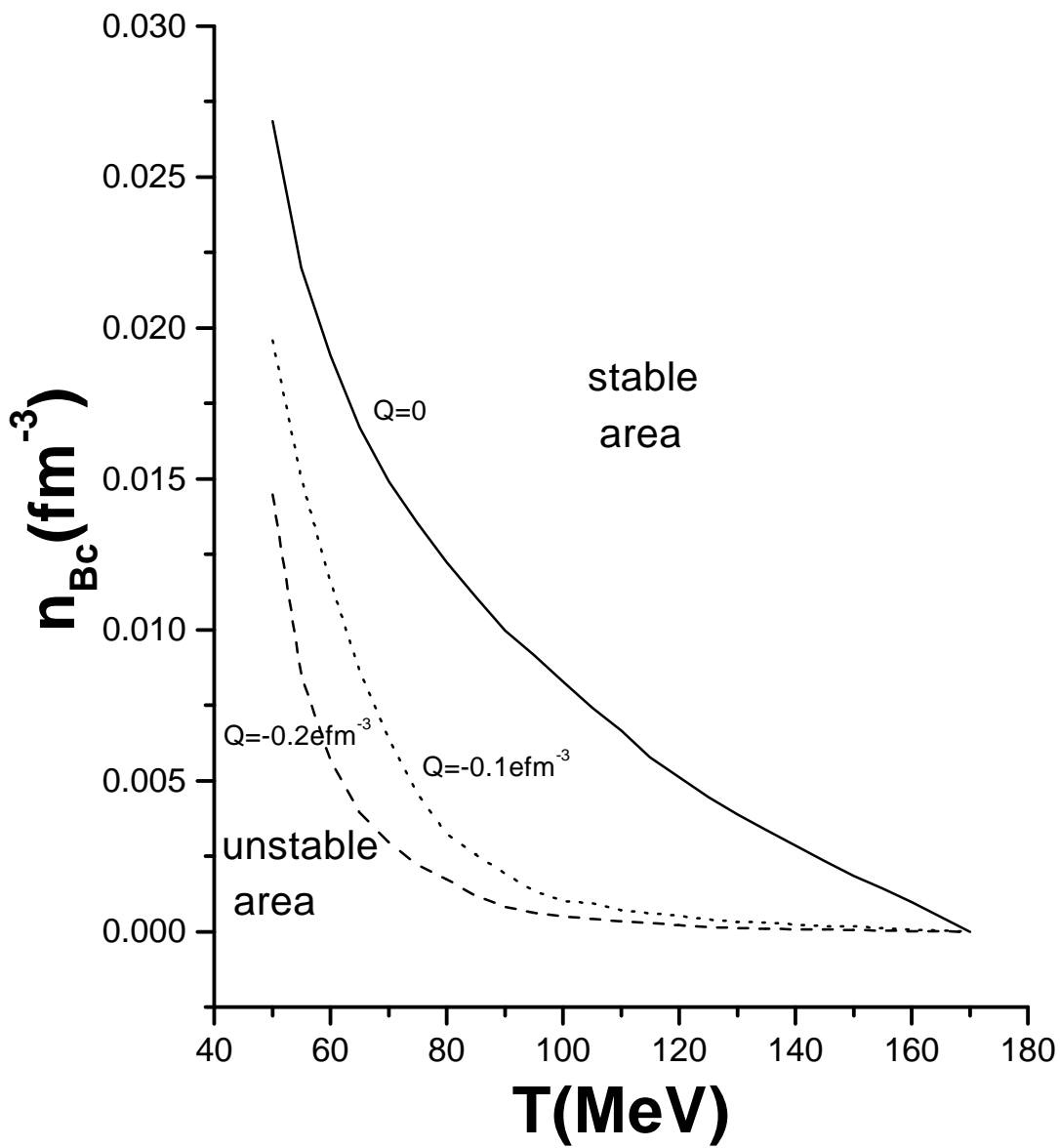
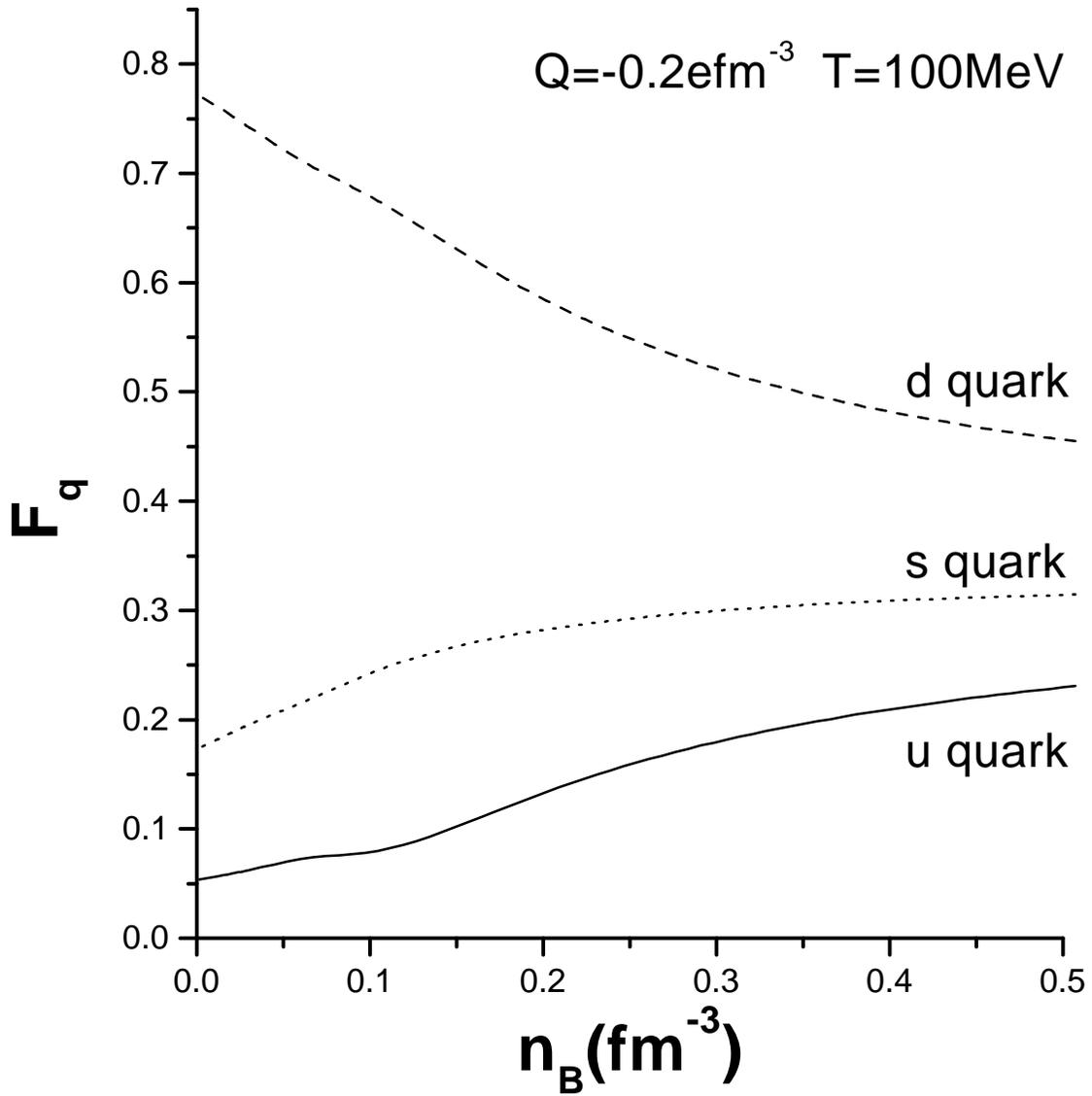
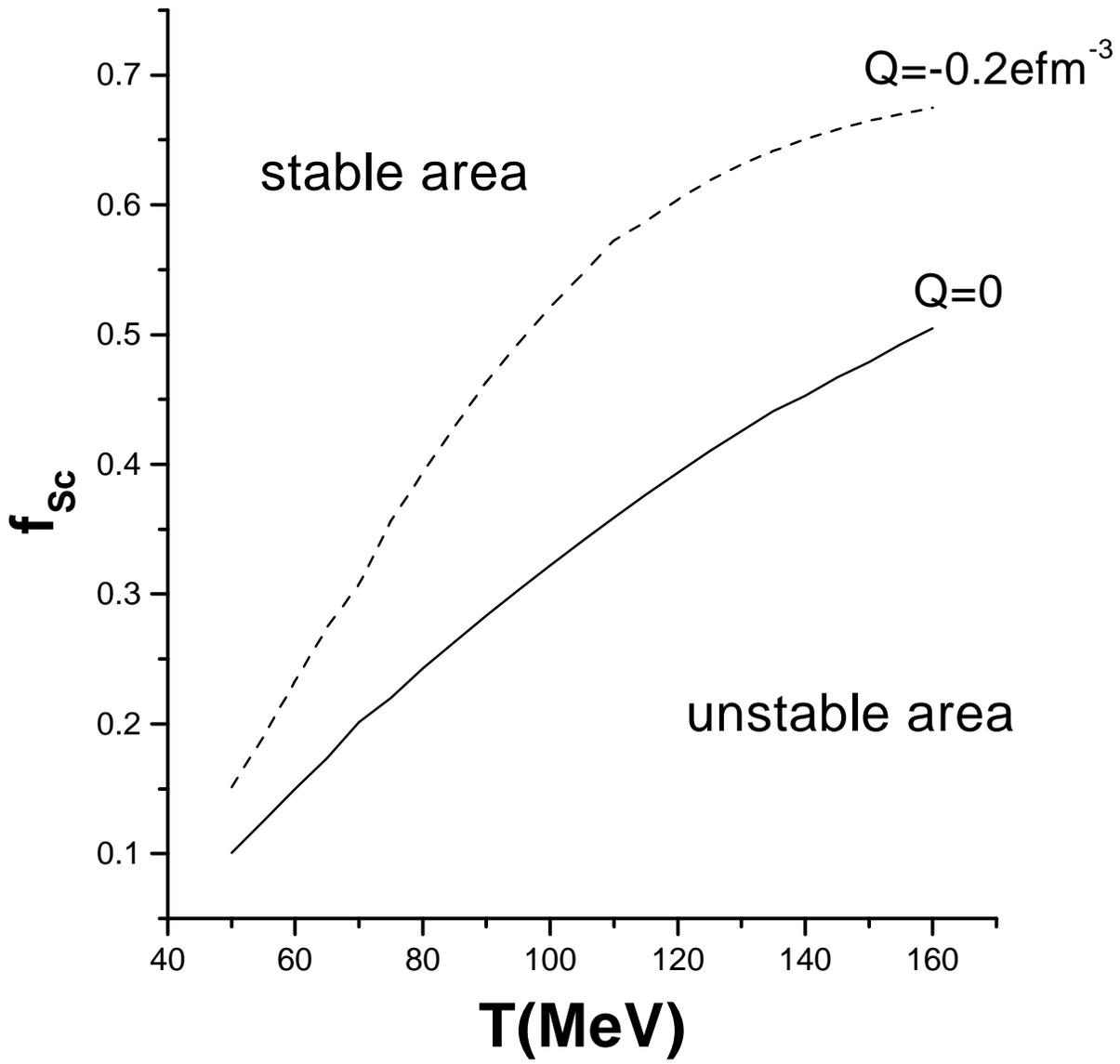
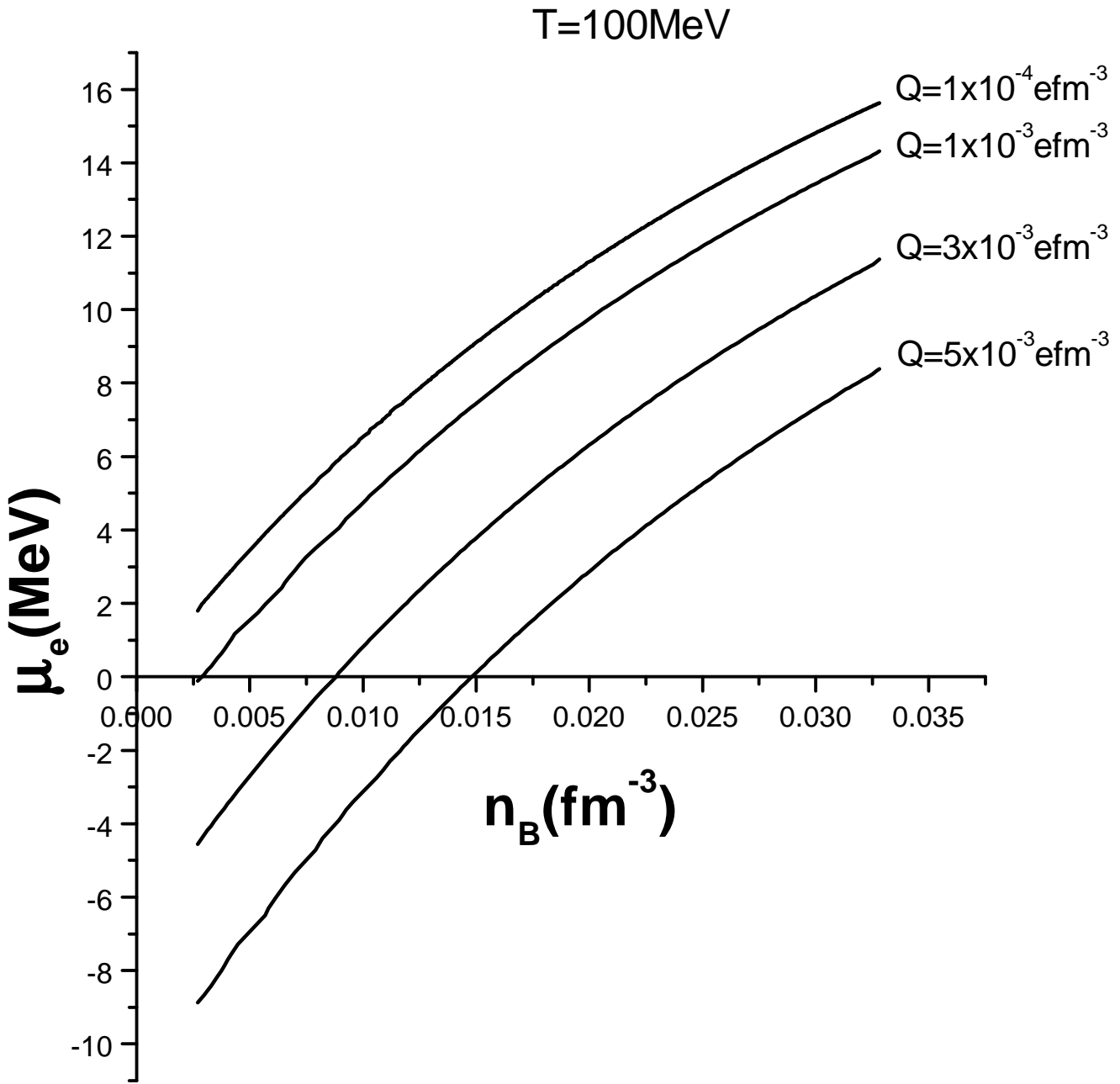
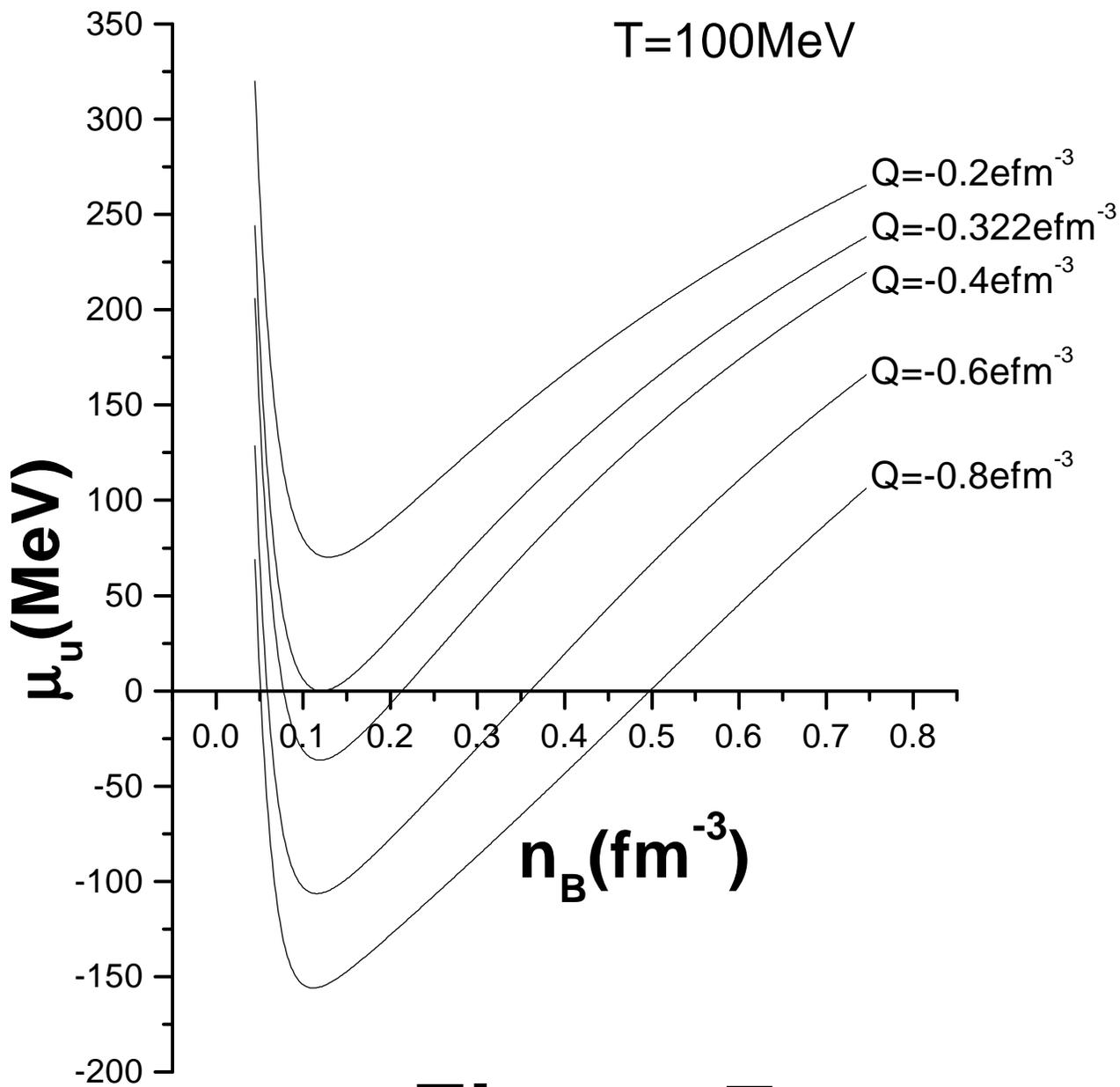
\end{document}